\documentclass[preprint,11pt,nofootinbib]{revtex4-1}
\usepackage{psfrag}
\usepackage{geometry} 
\usepackage{xcolor}
\usepackage{latexsym}
\usepackage{amsmath}
\usepackage{transparent}
 \usepackage{amscd}
 \usepackage{amsthm}
 \usepackage{amssymb}
\usepackage{graphicx}

\newcommand{\f}[2]{\frac{#1}{#2}}

\newcommand{\be}{\begin{equation}}

\newcommand{\ee}{\end{equation}}
\newcommand{\bea}{\begin{eqnarray}}
\newcommand{\eea}{\end{eqnarray}}

\newcommand{\nn}{\nonumber}

\def\cal#1{\mathcal{#1}}

\def\ptl{\partial}
\def\gp{g'}

\def\call{{\cal L}}

\def\Tr{\mathsf{Tr}}

\def\gev{~{\rm GeV}}

\def\eps{{\epsilon}}

\def\munu{{\mu \nu}}

\def\gp{g'}
\def\gpt{{\tilde g}^{\prime }}

\def\eps{{\epsilon}}

\def\Wt{\tilde{ {W}}}
\def\Bt{\tilde{ {B}}}

\def\s{s_\theta}

\def\gt{\tilde g}

\def\Zt{\tilde Z}
\def\vt{\tilde v}

\begin{document}

 \title{Rescaling mechanism and effective symmetry from the ideal cancellation of the $S$ parameter in custodial models}

\author{Donatello Dolce}  
\affiliation{University of Camerino, Piazza Cavour 19F, 62032, Italy. \\
 E-mail: \texttt{donatello.dolce@unicam.it}  
 }
\date{\today}
\begin{abstract}{We present a model independent analysis of custodial corrections to the $S$ parameter.  The negative contributions coming from direct corrections, i.e. the corrections associated to effects of new physics in the fermionic sector, can be used to eliminate unwanted positive oblique contributions to $S$. By means of such an ideal cancellation among oblique and direct corrections the electroweak physics can be made insensitive to custodial new physics. The Lagrangian analysis of the ideal cancellation reveals a possible sizable redefinition of the effective electroweak energy scale with respect to models with only oblique corrections. We then infer from general principles the full expression for the effective custodial operator responsible in the gauge sector for the contribution to $S$. We show that the ideal cancellation exactly eliminates all the custodial corrections, including those in the non-abelian and longitudinal terms. Indeed, provided redefinitions of the physical parameters, the standard model Lagrangian can be equivalently defined modulo custodial corrections in the gauge and fermionic sectors. Thus the ideal cancellation can be regarded as an effective symmetry. Finally we investigate the theoretical origin of this effective symmetry in terms of the effective gauge structure induced by custodial new physics.}  
\end{abstract}


\maketitle
\addcontentsline{toc}{section}{Introduction}

\section*{Introduction\label{sec:0}}

The  discovery of the Standard Model (SM) Higgs-like boson at LHC with mass $M_H \simeq 126 \gev$ represents a fundamental milestone to clarify the nature of the ElectroWeak (EW) gauge symmetry breaking and the origin of the elementary particle masses. Remarkably this seems to complete the SM picture. On the other hand it implies reconsiderations of some of the most investigated extensions of the Standard Model (SM) proposed to cure the unsatisfactory theoretical aspects of the Brout-Englert-Higgs mechanism (commonly known as the Higgs mechanism). In particular such a light SM Higgs boson implies fine-tunings or rescaling mechanisms to make the physics at the EW scale insensitive to possible new physics sectors. 

Effects of new physics at the $Z$-pole  are summarized by a set of EW parameters \cite{Beringer:1900zz}. The most relevant quantities parametrizing the EW observables are $T$, $U$ and $S$ \cite{Peskin:1991sw} or, in an alternative notation, $\eps_1$, $\eps_2$ and $\eps_3$ \cite{Altarelli:1993sz}.  The Peskin-Takeuchi parameters $T$ and $U$ can be generally made small or vanishing by requiring  custodial symmetric new physics, i.e. characterized by a symmetry breaking pattern $SU_L(2) \otimes SU_R(2) \rightarrow SU_D(2)$, where $SU_D(2)$ is the diagonal subgroup.   Custodial extensions of the SM  \cite{Agashe:2007mc,SekharChivukula:2009if,Mintakevich:2009wz,Cort:2013foa,Carena:2014ria} are protected from new physics corrections to the $T$ and $U$ precision parameters, but they are typically affected by large positive \emph{oblique} corrections to the $S$ parameter \cite{Barbieri:2004qk,Agashe:2007mc,Casagrande:2010si}. This create tensions with respect to the experimental bounds. 
The \emph{oblique} contributions to the EW parameters are associated to corrections to the  tree-level transverse vacuum polarization amplitudes of the SM gauge bosons.  However, if new physics is allowed in the gauge sector, corresponding new physics contributions can be assumed in the fermionic sector. In principle the contributions coming from the fermionic sector, known as  vertex or \emph{direct} corrections, can be as big as those in the gauge sector. 

In this paper we present a model independent analysis of the \emph{oblique} and \emph{direct} contributions to the $S$ parameter by introducing, within the framework of the SM of the EW interactions,  possible dimension six custodial operators in both the gauge and fermionic sectors \cite{BuchmŸller1986621,Han:2004az,Barbieri:2004qk,Agashe:2006at}. In particular we assume a SM Higgs sector. As we will show in sec.(\ref{EW:corrections}), \emph{direct} corrections give negative contributions to the $S$ parameter. They can be therefore fine-tuned with the \emph{oblique} contributions to conciliate the new physics corrections with the experimental bounds \cite{Anichini:1994xx,Cacciapaglia:2004rb,Casalbuoni:2005rs,Chivukula:2005xm,Casalbuoni:2007xn,Quiros:2013yaa}. We define \emph{ideal cancellation} the condition that exactly cancels each other out the oblique and direct corrections of new physics. 
 
 As known from extra-dimensional or moose models, the ideal cancellation of the $S$ parameter implies that new possible vectorial resonances are fermiophobic so that  their masses can be lower than expected in models with only oblique corrections, \cite{Chivukula:2005cc,Casalbuoni:2005rs,Accomando:2010fz}. For instance, in extra-dimensional theories (or moose models), the ideal delocalization means that  the fermionic Kaluza-Klein towers in the bulk (or along a moose) are proportional to the  delocalization of the vectorial Kaluza-Klein tower, in particular to the $W$ delocalization. Since the new vectorial resonances, such as $Z'$ and $W'$, are orthogonal to the SM $Z$ and $W$,  upon redefinitions of the physical quantities the ideal cancellation implies that the coupling of these new vectorial to the SM fermions are strongly suppressed.  Similarly, these considerations can be extended to technicolor and composite models.  

The standard analysis of oblique and direct corrections is based on the evaluation of the so-called $\Delta$ parameters. 
This method must be however carefully used. Being based on the ratio of the vector boson masses $M_W/M_Z$ it does not reveal possible redefinition of the effective EW scale, i.e. of the Fermi coupling. Indeed the direct contributions can lead to redefinitions of effective vector boson masses, Yukawa couplings and parameters in the Higgs sector, with respect to the case in which only oblique corrections are assumed. In  sec.(\ref{lagrang:analysis}), to see these rescalings and the exact condition which relates the oblique and direct corrections in the ideal cancellation, we will perform the explicit  Lagrangian analysis through the redefinition of the SM parameters.  Both the extended parameters region allowed by the ideal cancellation and the redefinition of the EW scale have important consequences in model building: physics at the $Z$-pole can be made insensitive to custodial new physics sectors and the bounds of new physics can be in principle lowered.  
Furthermore, the operator responsible for the direct custodial corrections investigated in this paper plays an important role in curing the tensions of some extensions of the SM in reproducing the correct Top quark mass and $Z b \bar b$ coupling, \cite{Agashe:2006at,Casagrande:2008hr,Cui:2009dv}. 
 
 In sec.(\ref{effective:sym}) we will infer from general principles, namely the requirement of custodial invariance and the elimination of heavy modes from the Equations of Motion (EoM) in the derivation of the effective theory, the correct form of the operator $\hat O_{WB}$ responsible in the gauge sector for the contributions to the $S$ parameter. In this way we will find that, besides the exact elimination of the corrections of new physics in the transverse components of the vector bosons, the ideal cancellation exactly eliminates also all the spurious corrections in the longitudinal and non-abelian terms of the effective Lagrangian. Thus, as long as new physics can be neglected in the EoM, the SM Lagrangian turns out to be equivalently defined modulo custodial oblique and direct operators (the concept of ideal cancellation can be also extended to the Higgs sector, \cite{Quiros:2013yaa}), provided redefinitions of the physical parameters.  
 In analogy with the case of custodial symmetry protecting $T$ and $U$,  the ideal cancellations can be thought of as an ``effective symmetry'' (i.e. a symmetry valid at low energy) of the SM Lagrangian protecting the $S$ parameter. 
 
In sec.(\ref{spurious}) we will investigate the possible theoretical origin of the ideal cancellation. 
It is known that in extra-dimensional, holographic, moose, composite and technicolor models the ideal cancellation is related to a fine-tuning of the direct coupling of SM fermions to the new physics sector.  We will find that the origin of the custodial oblique and direct corrections can be associated to  modifications of the covariant derivatives in the gauge and fermionic sectors. At the level of covariant derivative, the ideal cancellation means that these modifications must be the same in both the gauge and fermionic sectors. This suggests that the ideal cancellation has a possible interpretation in terms of the ``universality'' of gauge structure among the effective vectorial and fermionic sectors. 

\section{Electroweak corrections of custodial new physics}\label{EW:corrections}

The contributions of new physics in extensions of the SM of EW interactions are mainly summarized by three EW parameters $T$, $U$ and $S$ \cite{Peskin:1991sw}, or alternatively the $\eps_1$, $\eps_2$ and $\eps_3$ parameters \cite{Altarelli:1993sz}. These corrections are inferred from the effective Lagrangian through three $\Delta$ parameters: $\Delta r_W$ is related to the $M_W$ and $M_Z$ ratio, $\Delta_\rho$ is related to the neutral and charged current interactions, and $\Delta_k$ takes into account the difference between effective value of the Weinberg angle, $\tilde \theta$, and its physical value $\theta$:
\begin{eqnarray} 
\call^{fermions}_Z &=& - \frac{e}{s_{\theta}c_{\theta}}\left(1 + \frac{\Delta \rho^{}}{2}\right)[J_{3L}^\mu - s_{\theta}^2 (1 + \Delta k^{})J_{em}^\mu] Z_{_{}\mu}, \label{lagrZ:Delta:SM} \nn \\
s_{\tilde{\theta}_{}}^2 &=& (1 + \Delta k^{})s_{\theta}^2, \label{stheta:Deltak:SM} \nn \\
\left(1 - \frac{M_{W_{}}^2}{M_{Z_{}}^2}\right)\frac{M_{W_{}}^2}{M_{Z_{}}^2} &=& \frac{\pi \alpha(M_Z)}{\sqrt{2}G_F M_{Z_{}}^2(1 - \Delta {r_W}^{})}, \label{massebos:Deltar:SM} 
\end{eqnarray} 

 We define the following reduced EW parameters  \cite{Altarelli:1993sz,Beringer:1900zz},
\begin{eqnarray}
 \hat T &=& \Delta\rho\,,\nn\\
 \hat U &=& c_\theta^2 \Delta\rho+\f{\s^2}{c_{2\theta}}\Delta r_W-2 \s^2\Delta k\,,\nn\\
\hat S &=& c_\theta^2 \Delta\rho+c_{2\theta}\Delta k \,.
\label{epsdef}
\end{eqnarray}
In an alternative notation these parameters correspond respectively to the $\eps_1$, $\eps_2$ and $\eps_3$ parameters \cite{Altarelli:1993sz}.  

In custodial models the parameters $T$ and $U$ are protected from new physics contributions. Notice, however, that besides the contribution of $ T$,  $ U$ and $ S$, the reduced parameters defined in (\ref{epsdef}) may have contributions from radiative corrections as well as from higher derivative corrections $V$, $W$, $X$, $Y$  and $Z$.  A non-SM Higgs sector typically implies corrections to $T$ and $S$, however in this paper we will work by assuming a SM Higgs sector, i.e. that new physics does not contribute to the SM particle masses. In general, higher derivative contributions of the same type can be consistently neglected with respect to the lower derivative ones as they are suppressed by factors $(M_W / \Lambda)^2$ where $\Lambda$ is the new physics scale. The remaining lower derivative relevant parameters  $W$ and $Y$  preserve the custodial symmetry. They can be relevant in theories with new or composite vector bosons \cite{Barbieri:2004qk}.  Nevertheless, we will work under the assumption of ideal cancellation and this implies that new or composite resonances are fermiophobic so that their effective contributions to $W$ and $Y$ are suppressed or vanishing \cite{Chivukula:2005cc}. The scope of this paper is to investigate the $S$ parameter. The detailed Lagrangian analysis of the other EW precision parameters will be given in future publications and our analysis will be at tree-level. Thus the reduced EW parameters (\ref{epsdef}) are related  to the ordinary EW parameters by the relations: $ U = - 4 s^2_\theta \hat U / \alpha \simeq  -119 ~ \hat U$, $ T = \hat T / \alpha = 129~  \hat T $ and $ S =  4 s^2_\theta \hat S / \alpha \sim  119 ~ \hat S$, \cite{Barbieri:2004qk}. The $S$ related values assumed in this paper are $\eps_3 = (4.8 \pm 1.0)\times 10^{-3}$ (experimental bounds) and $\hat S = (0.0 \pm 1.3)\times 10^{-3}$ (global fit with a light Higgs) \cite{Barbieri:2004qk}.

\subsection{Oblique corrections}

The so-called oblique corrections to the EW parameters are directly inferred from the transverse components of vectorial fields. If new vectorial resonancies are sufficiently heavy with respect to the mass of the external particles in Feynman diagrams the longitudinal part of the two point functions are suppressed by factors $(m_f/m_V)^2$. 
The expansion in $p^2$ of the tree-level transverse vacuum polarization amplitudes $\Pi_{V_1 V_2}(p^2) \simeq \Pi_{V_1 V_2}(0) + \dot\Pi_{V_1 V_2}(0)p^2 + \mathcal{O}(p^2)$, where $V_1$ and $V_2$ are generic SM gauge bosons, leads to the following definitions of the oblique EW corrections \cite{Peskin:1991sw}:
\begin{eqnarray}\label{eps:par:def:vacuum}
 \hat T_{Obl} &=& 
- \frac{1}{2} \frac{\Pi_{W W}(0) - \Pi_{\pm \pm}(0)}{\Pi_{W W}(0)}~,\nn \\
 \hat U_{Obl} &=& -  (\dot{\Pi}_{W W}(0) - \dot{\Pi}_{\pm \pm}(0))~,\nn \\
\hat S_{Obl} &=&  \frac{g}{g'} \dot{\Pi}(0)_{W Y}~,
\end{eqnarray}

Custodial extensions of the SM, such as custodial holographic, extra-dimensional, moose, composite-Higgs, Little Higgs, and technicolor models, are typically affected by positive contributions to the  $S$ parameter, whereas $T$ and $U$ are protected by the custodial symmetry. 
 In the literature, e.g. \cite{BuchmŸller1986621,Han:2004az,Barbieri:2004qk,Fichet:2013ola,Contino:2010rs}, the  oblique contributions to $S$ are commonly associated to the operator
 \be 
 \mathcal O_{WB} =  (U \sigma^a U^\dagger) F^a_\munu(\Wt) \Bt^\munu \leadsto  - \frac 1 2  \frac{\gpt}{\gt}\hat S_{Obl}   F^3_\munu(\Wt) \Bt^\munu\label{op:S:Obl}\,.
 \ee 
 where $\hat S_{Obl}$ is positive, $U$ is the chiral field transforming under $SU_L(2) \otimes SU_R(2)$ as $U \rightarrow L U R^\dagger$, $F_\munu^a(\Wt) = \Wt^a_\munu +  \gt \eps^{abc}\Wt_\mu^b \Wt_\nu^c$ and $ \Wt^a_\munu = \ptl_\mu  \Wt^a_\nu - \ptl_\nu  \Wt^a_\mu$. This operator generates anomalous trilinear vector boson couplings \cite{Han:2004az}.  However, in sec.(\ref{spurious}) we will infer from general principles the correct effective operator associated to custodial corrections, namely $\hat {\mathcal O}_{WB}$, see (\ref{correct:oblique:operator}). This indeed differs from ${\mathcal O}_{WB}$ in the non-abelian terms. Since in this and in the next section we only consider the transverse components of the vector bosons such a difference in the operator $\mathcal O_{WB}$ can be neglect for the moment. 
Thus, in momentum space, the effective Lagrangian encoding oblique corrections of custodial new physics  is  
\begin{eqnarray}
\call^{Eff}_{Obl}(p^2, \vt)
= \call^{SM}(p^2, \vt) 
+ p^2 \frac{\gpt}{\gt}\hat S_{Obl}   \Wt^{3}_{\mu} \Bt^\mu \,.
 \label{S:obl:act}
\end{eqnarray}

The effective electroweak scale $\tilde{v} $ and the effective Weinberg angle $\tilde \theta$  s.t. $\tan{\tilde \theta} =  \gpt / \gt$, determine the effective  vector boson mass spectrum associated to (\ref{S:obl:act}): 
\begin{eqnarray}\label{Obl:vect:mass}
\tilde M_W = \gt \frac{\vt}{2} ~~,~~~ \tilde M_Z = \sqrt{\gt^2 + {\tilde{g'}}^2}  \frac{\vt}{2}
\end{eqnarray} 
We assume a SM Yukawa and Higgs sector so that the fermion masses are determined by the Yukawa couplings and are proportional to the EW scale  
\be
\tilde m_f = \frac{1}{\sqrt 2} \tilde \lambda_f \vt \,,
\ee
and the Higgs boson mass is determined by the SM Higgs potential parameter $\tilde \lambda$ and the  EW scale
\be
\tilde M_H = \sqrt{2 \tilde \lambda} \vt \,.
\ee 
In other words we are assuming that the SM masses are generated by the SM Higgs mechanism, i.e  new physics does not contribute to the SM mass spectrum. This means that the contribution of new physics to the  vacuum polarization amplitudes is supposed to be zero at leading order in $p^2$, i.e. $\Pi_{V_1 V_2}(0)\equiv 0$, and that the Higgs field has no couplings to the new physics sector. 
In case of alternative gauge symmetry breaking mechanisms the parameters $\tilde \lambda_f$ and $\tilde \lambda$ are model dependent (with negligible higher order corrections associated to the effective EW parameters $\gt, \gpt$ and $\tilde{v}$). As exemplification, we will mention later some model dependent cases.

\subsection{Direct corrections}\label{Direct:corrections}

Besides the oblique corrections it is possible to introduce corrections affecting the vertices with the fermions. This are the so-called direct or vertex corrections. Indeed, the parameters (\ref{epsdef}) encode general corrections  to neutral and charged currents. Thus they are affected by  direct couplings of  the SM fermions to new physics, FIG.(1). 
\begin{figure}[htbp]
\begin{center}\label{Fdiagram:obl:dir}
\def\svgwidth{\columnwidth}
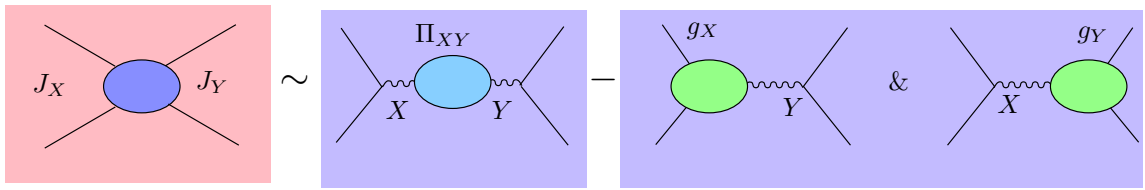
\caption{A diagrammatic representation of the oblique and direct contributions to a current-current interaction.} 
\end{center}
\end{figure}

A possible universal direct contribution to the $S$ parameter allowed by the custodial symmetry is given by the operator \cite{Han:2004az,Agashe:2006at,Casalbuoni:2005rs,Casalbuoni:2007xn}
\begin{eqnarray}
 \mathcal C_2 =  \Tr [\bar\psi_L \gamma_\mu  (i D^\mu U) U^\dagger \psi_L] &\leadsto&  c_2 \gamma^\mu \bar\psi_L \left( \gt  \Wt^a_{ \mu}\frac{\sigma^a}{2} -   \gpt \Bt_{ \mu} \frac{\sigma^3}{2} \right)\psi_L \nn \\
&=& c_2 \frac{\tilde e}{\sin \tilde \theta \cos \tilde \theta } \Zt_\mu  \bar\psi_L \gamma^\mu \frac{\sigma^3}{2} \psi_L  + c_2 \frac{1}{\sin \tilde \theta \sqrt 2} \Wt_\mu^- \bar\psi_{Ld} \gamma^\mu \psi_{Lu}  
  \label{S:dir:operator}
\end{eqnarray}
 The sign of $c_2$ is determined by the kinetic term and it is positive.  $ \mathcal C_2$  is for instance associated to the universal delocalization of left-handed fermions in custodial extra-dimensional models or holographic models \cite{Cacciapaglia:2004rb,Casalbuoni:2007xn} and, through dimensional deconstruction \cite{Arkani-Hamed:2001ca}, to direct couplings of standard model fermions to the non-standard gauge symmetries of moose models \cite{Casalbuoni:2005rs,Chivukula:2005xm} or similar strongly-interacting and composite models \cite{Anichini:1994xx,Accomando:2012yg,Contino:2010rs}. Furthermore, as noted in \cite{Agashe:2006at}, this operator applied to the third quark family is important to cure the $Z b \bar b$ coupling in emergent EWSB models. 
 
When both oblique and direct custodial corrections are considered the effective Lagrangian is 
\be\label{LEFF:OBL:DIR}
\call^{Eff}_{Tot}(p^2,\vt) 
=  \call^{SM}(p^2,\vt) 
+ \mathcal O_{WB} + \mathcal C_2\,.
\ee
Next we will investigate the EW corrections of this effective custodial Lagrangian. 

 \section{Standard analysis of the $S$ corrections}

The standard analysis of the EW corrections associated to the effective Lagrangian (\ref{LEFF:OBL:DIR})  passes through the evaluation of the $\Delta$ parameters (\ref{massebos:Deltar:SM}).  
 The explicit calculation of the both oblique and direct contributions yields 
  \begin{eqnarray}\label{eps:custodial}
  \Delta \rho \simeq 0~;~~ 
  \Delta r_W \simeq - 2(c_2 -  \hat S_{Obl})~;~~ 
  \Delta k \simeq -  \frac{c_2 -  \hat S_{Obl}} { \cos 2 \tilde \theta } ~.
  \end{eqnarray}

  From this we find that $\hat U \sim 0$ and $\hat T \sim 0$ as the oblique and direct contributions of $\Delta r_W$ and $\Delta k$ cancel each other out in (\ref{epsdef}). This result was expected from the fact that both oblique and direct contributions, (\ref{op:S:Obl}) and (\ref{S:dir:operator}), preserve the custodial symmetry, and $\hat U$ and $\hat T$, as well as the $\rho$ parameter, are protected by the custodial symmetry. 
  
  The $\hat S$ parameter is in general non vanishing in custodial models. From (\ref{epsdef}) and (\ref{eps:custodial})  we find
  \be
\hat  S \simeq \hat S_{Obl} - c_2 \,.\label{i-canc:c2:std}
  \ee
  so that the direct contribution to the $\hat S$ parameter is negative: $\hat  S_{Dir} \simeq - c_2 $.

  In the case of fine-tuning  between oblique and direct corrections
  \be\label{ideal:canc:approx}
   c_2  \simeq \hat S_{Obl}   \,,
   \ee  
  all the $\Delta$ parameters are suppressed. Thus the $\hat S \sim 0$ parameter  is also protected from corrections of new physics. Summarizing,  the three EW corrections are vanishing  
  \be
  \hat U \simeq 0 ~;~~\hat T \simeq 0 ~;~~\hat S \simeq 0 ~.
  \ee
  
  We define \emph{ideal cancellation} the condition that exactly cancel the oblique and direct contributions to the $\hat S$ parameter. 
   In custodial extra-dimensional or moose models, this corresponds to the ``ideal delocalization'' of left-handed fermions along the extra-dimensional bulk \cite{Cacciapaglia:2004rb,Casalbuoni:2007xn,Quiros:2013yaa} or along a moose \cite{Casalbuoni:2005rs,Chivukula:2005xm}, or other ideal couplings of fermions to custodial extensions of the EW gauge group in composite-Higgs, little-Higgs and technicolor models \cite{Anichini:1994xx,Accomando:2012yg}. 
   
   In extra-dimensional models this requires that the left-handed fermions bulk profile is related to the $W$ profile.   According to the bulk EoM, the two profiles at zero order in $p^2$ behave as $f_L(z)\sim z^{2(2+c) \beta_L}$ and $h_W(z) \sim \alpha_W + \beta_W z^2$ respectively, where: $z=e^{ky}/k$ is the conformal parametrization of the warped extra-dimension $y$; $\beta_L$, $\alpha_W$ and $\beta_W$ are constants determined by the boundary conditions; $c= M_{bulk} / k$ is the ratio between the fermion bulk mass and the curvature of the AdS metric. Thus the condition of ideal delocalization $h_W (z) \sim (k z)^{-3} f^2_L(z)$  is of difficult realization:  it is necessary to have a flat fermionic profile, namely $c= - 1 /2$, and to arrange the boundary conditions in such a way that $\alpha_W\sim 0$ and $\beta_W \sim \beta^2_L/k^3$.     
   Notice that, concerning the fine-tuning (\ref{ideal:canc:approx}),  if the corrections in the gauge and in the fermionic sectors come from the same new physics they can give, in principle, contributions of the same order in the effective Lagrangian.\footnote{As the ideal cancellation related corrections to the gauge sector with those of the fermionic sector, it would be particularly interesting to study this ideal cancellation in supersymmetric models (or in terms of the BRST symmetry).
}
   
In such a standard analysis of the EW corrections the effect  of possible custodial direct corrections to the vector boson mass spectrum (such that $\rho = 1$) is typically neglected, i.e. the mass spectrum associated to the effective Lagrangian (\ref{LEFF:OBL:DIR}) is typically assumed to be (\ref{Obl:vect:mass}) \cite{Anichini:1994xx,Cacciapaglia:2004rb,Casalbuoni:2005rs,Chivukula:2005xm,Casalbuoni:2007xn} regardless direct corrections. This is because  the standard analysis of the EW parameters encodes only the oblique corrections written in terms of the gauge bosons mass ratio \cite{Beringer:1900zz,Altarelli:1993sz}.  
In the next section we will see that the analysis of the EW direct corrections performed at the Lagrangian level reveals a rescaling of the effective value $\vt$ associated to the mass spectrum (\ref{Obl:vect:mass}) of models with only oblique corrections with respect to the physical EW energy $v$. We will find that this can lead to sizable corrections in the case of ideal cancellation. 

In sec.(\ref{effective:sym}), we will find that the assumption of ideal cancellations represents an effective symmetry of the SM lagrangian. That is, the gauge and fermionic sectors of the SM Lagrangian can be defined modulo oblique and direct terms, $\hat{\mathcal O}_{WB}$ and $\mathcal C_2$ respectively. Such an ``effective symmetry''  is indeed exact  as soon as we consider the general form of the oblique operator $\hat{\mathcal O}_{WB}$, (\ref{correct:oblique:operator}) derived through the elimination of the heavy d.o.f. from the EoM.   

 \section{Lagrangian analysis of the $S$ corrections}\label{lagrang:analysis}
 
 To perform the Lagrangian analysis of the EW corrections we explicitly act in the effective Lagrangian containing both the oblique and direct corrections (\ref{LEFF:OBL:DIR}) with redefinitions of fields and couplings. In momentum space, neglecting for the moment the longitudinal and the non-abelian terms, (\ref{LEFF:OBL:DIR}) is
 \begin{eqnarray}\label{LEFF:OBL:DIR:expl}
\call^{Eff}_{Tot} (p^2,\vt) &=&   \frac {p^2} 2 \Wt^{a \mu} \Wt^a_{ \mu}  + \frac {p^2} 2  \Bt^\mu \Bt_\mu  + {p^2}    \frac \gpt \gt \hat S_{Obl} 
\Wt_\mu^3 \Bt^\mu  - 
\frac{\vt^2}{2} \text{Tr}\left[\gt \Wt^a \frac{\sigma^a}{2} - \gpt \Bt \frac{\sigma^3}{2}\right]^2   
\nn  \\ &+&
\bar\psi_L \gamma^\mu \left(p_\mu -  \gt \Wt_\mu^a \frac{\sigma^a}{2} -  \gpt \frac{B-L}{2} \Bt_\mu \right) \psi_L \nn \\ &+&
\bar\psi_R \gamma^\mu \left(p_\mu -  \gpt \Bt_\mu \frac{\sigma^3}{2} -  \gpt \frac{B-L}{2} \Bt_\mu \right) \psi_R \nn \\&-& 
c_2 \gamma^\mu \bar\psi_L \left(- \gt  \Wt^a_{ \mu}\frac{\sigma^a}{2} +   \gpt \Bt_{ \mu} \frac{\sigma^3}{2} \right)\psi_L +  \call^{SM}_{Higgs} +   \call^{SM}_{Yukawa}\,.
\end{eqnarray}

 We apply the following redefinitions of the gauge fields and couplings,  
\be\label {redef:S:canc}
\Wt^a_\mu \frac{\sigma^a}{2} \rightarrow W^a_\mu \frac{\sigma^a}{2} - \hat S_{Obl} \frac \gpt \gt  \Bt_\mu \frac{\sigma^3}{2}~,~~~\gt \rightarrow  \frac{g}{1-c_2} \,.
\ee
This diagonalizes the kinetic gauge terms. Besides this, we normalize the $\Bt_\mu$ kinetic term with the (unphysical) redefinitions $
\Bt_\mu \rightarrow {B_\mu}/\sqrt{1 +  \frac{3}{2} \frac{{\gt}^{\prime 2}}{\gt^2} \hat S^2_{Obl}}$ and $\gpt \rightarrow \gp \sqrt{1 +  \frac{3}{2} \frac{{\gt}^{\prime 2}}{\gt^2} \hat S^2_{Obl}}
$.
The effect of these transformations is to transfer the oblique corrections to the fermionic sector and to the vector boson mass term
\begin{eqnarray}
\call^{Eff}_{Tot}(p^2,\vt) &=&   \frac {p^2} 2 W^{a \mu} W^a_{ \mu}  + \frac {p^2} 2  B^\mu B_\mu 
+ \frac{\vt^2}{2} \text{Tr}\left[\frac{g}{ 1 - c_2}  W_\mu^a  \frac{\sigma^a}{2} - g' (1 + \hat S_{Obl} ) B_\mu  \frac{\sigma^3}{2} \right]^2  \nn  \\
&+& \bar\psi_L \gamma^\mu \left[p_\mu - g  W^a_\mu \frac{\sigma^a}{2} -  g' \left( {c_2} - \hat S_{Obl}+ \hat S_{Obl} c_2 \right) \frac{\sigma^3}{2}  B_\mu -  g'  \frac{B-L}{2} B_\mu\right] \psi_L \nn \\
&+&\bar\psi_R \gamma^\mu \left(p_\mu -  g' B_\mu \frac{\sigma^3}{2} -  g' \frac{B-L}{2} B_\mu \right) \psi_R +  \call^{SM}_{Higgs} +  \call^{SM}_{Yukawa} \,.
\end{eqnarray}

Thus we find that the ideal cancellation among oblique and direct corrections in the custodial effective Lagrangian is given by the following exact condition
\be\label{ideal:canc}
\boxed{c_2 \equiv  \frac{\hat S_{Obl}}{1 + \hat S_{Obl}}}\,,
\ee
so that $\hat S \equiv 0 $, $ \hat U \equiv 0$ and $\hat T \equiv 0$. 
This result is in agreement with that obtained at the first order in $\hat S_{Obl}$ through the standard analysis (\ref{i-canc:c2:std}, \ref{ideal:canc:approx}).  Indeed the condition (\ref{ideal:canc}) cancels the anomalous term in the left-handed neutral current. Furthermore it implies a factorization of the corrections in the vector boson mass term 
\begin{eqnarray}
\call^{Eff}_{Tot}(p^2,\vt) &=& \frac {p^2} 2 W^{a \mu} W^a_{ \mu}  + \frac {p^2} 2  B^\mu B_\mu  +   (1+ \hat S_{Obl})^2 \frac{\vt^2}{2}   \text{Tr}\left[g W_\mu^a \frac{\sigma^a}{2} - g'  B_\mu \frac{\sigma^3}{2} \right]^2    \nn  \\
&+& \bar\psi_L \gamma^\mu \left(p_\mu -  g \Wt_\mu^a \frac{\sigma^a}{2}  -  \gp \frac{B-L}{2} \Bt_\mu\right) \psi_L \nn \\
&+&\bar\psi_R \gamma^\mu \left(p_\mu -  \gp \Bt_\mu \frac{\sigma^3}{2} -  \gp \frac{B-L}{2} \Bt_\mu \right) \psi_R 
+  \call^{SM}_{Higgs} +  \call^{SM}_{Yukawa} 
\end{eqnarray}

\begin{figure}[tbp]
\centering 
\includegraphics[width=0.65\textwidth]{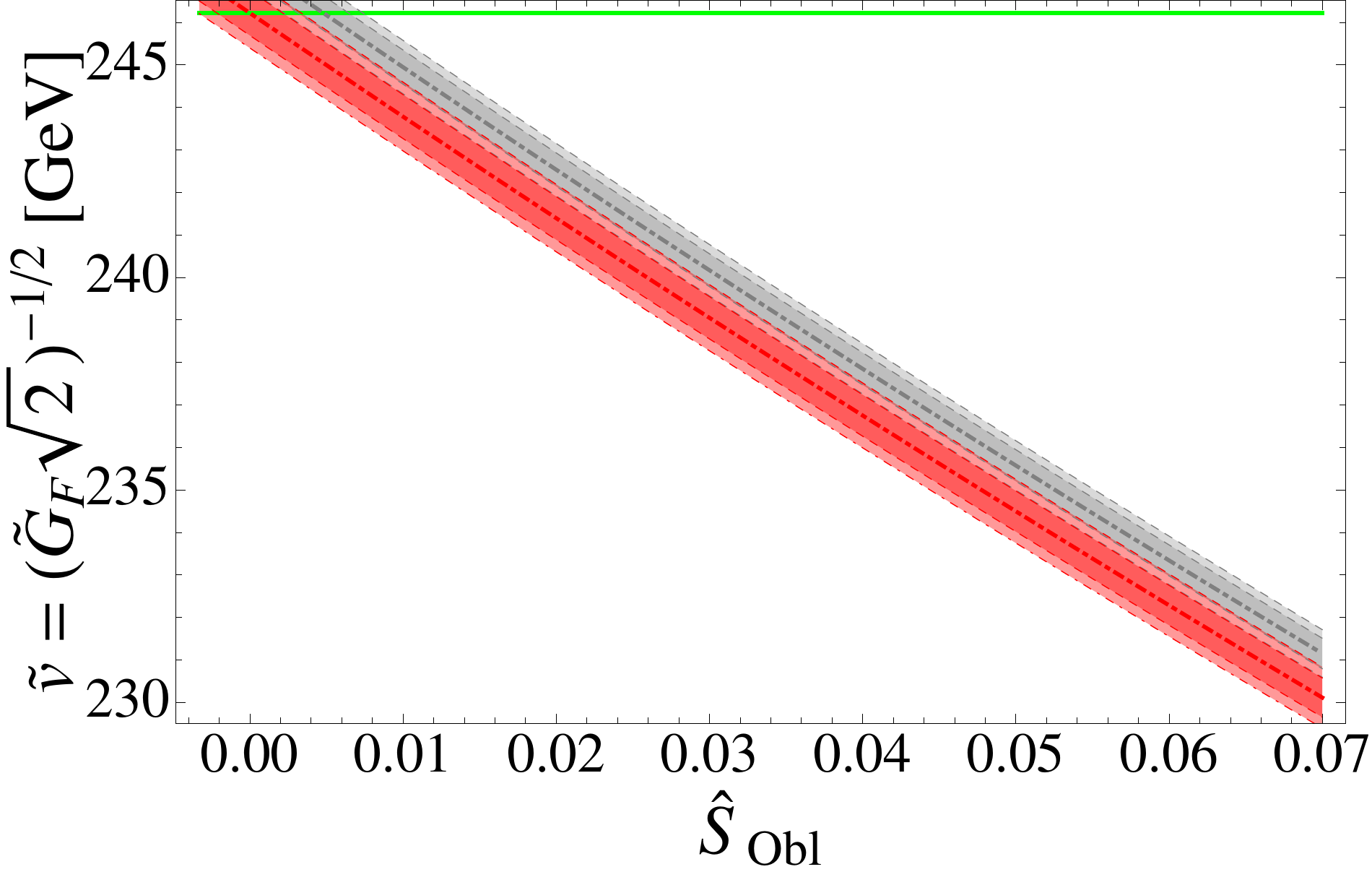} 
\caption{\label{fig:1} 
90\% and 99\% CL contour plot for the rescaling of the effective EW energy scale (and Fermi coupling) resulting from the ideal cancellation in the parameter $\hat S$ (red region) and $\eps_3$ (gray region). $\vt \equiv (\tilde G_F \sqrt{2})^{-\frac{1}{2}}= v / (1 + \hat S_{Obl})$ is the effective value calculated in custodial models by only  considering oblique contributions, $v \equiv ( G_F \sqrt{2})^{-\frac{1}{2}} = 246.22 \gev$ (green line) is the physical value obtained by also considering direct corrections.}
\end{figure}

Through the ideal cancellations the oblique and direct corrections are completely cancelled and the effective Lagrangian (\ref{LEFF:OBL:DIR:expl}), i.e. (\ref{LEFF:OBL:DIR}), is reduced to the SM Lagrangian 
\begin{eqnarray}
\boxed{\call^{Eff}_{Tot}(p^2,\tilde{v}) = \call^{SM}(p^2,\vt) + \hat {\mathcal O}_{WB} + \mathcal C_2 \equiv \call^{SM}(p^2,{v})}\,,
\end{eqnarray}
 with the following redefinition of the effective EW scale $\vt = \frac{1}{\sqrt{\tilde G_F} \sqrt{2}}$, i.e. of the effective Fermi coupling $\tilde G_F$, see FIG.\ref{fig:1},
\be\label{vev:rescaling:exact}
\boxed{\vt \rightarrow \frac{v}{1+ \hat S_{Obl}} = {v}{(1- c_2)}}\,.
\ee

\begin{figure}[tbp]
\centering 
\includegraphics[width=0.49\textwidth]{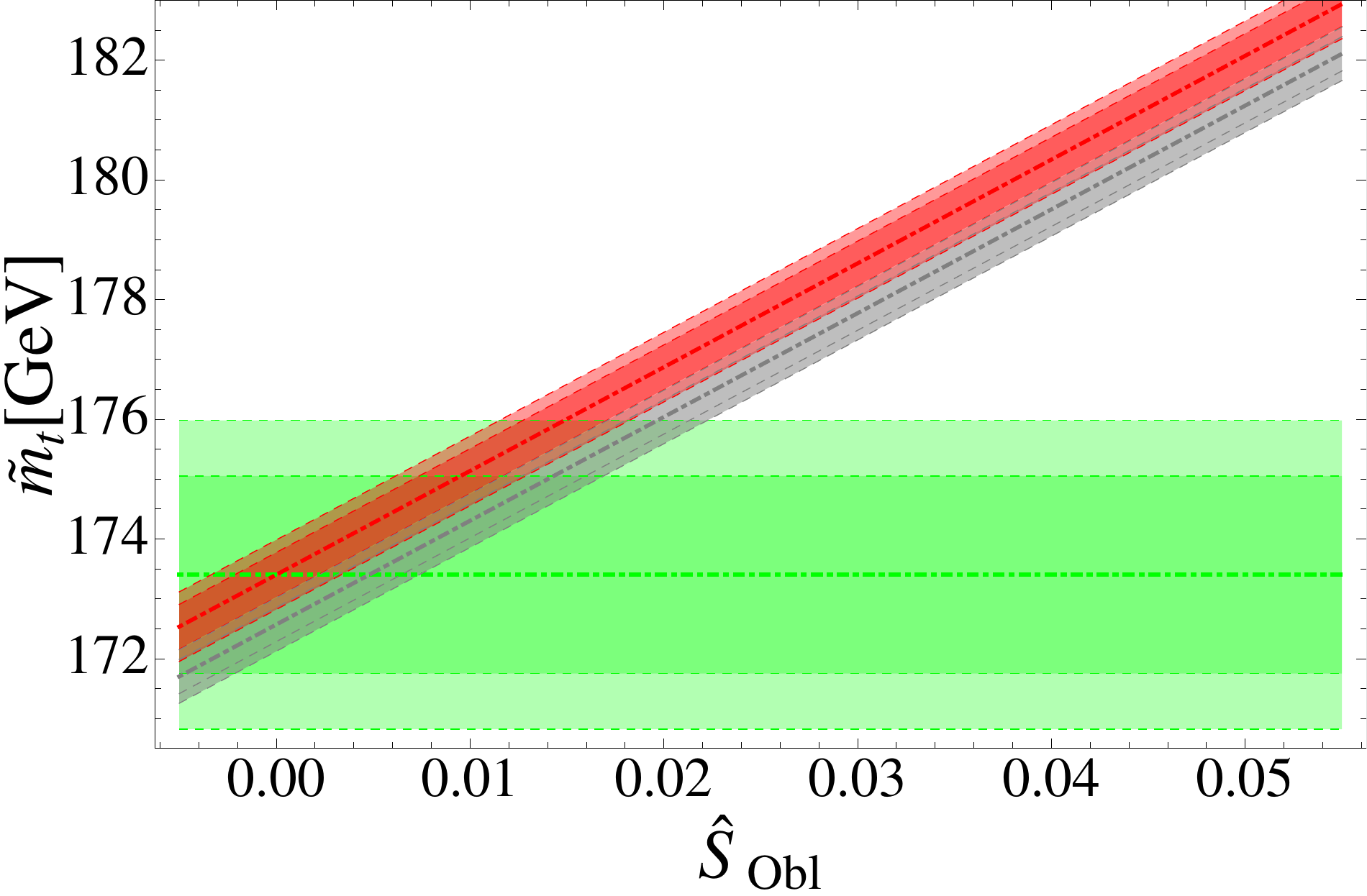} 
\includegraphics[width=0.49\textwidth]{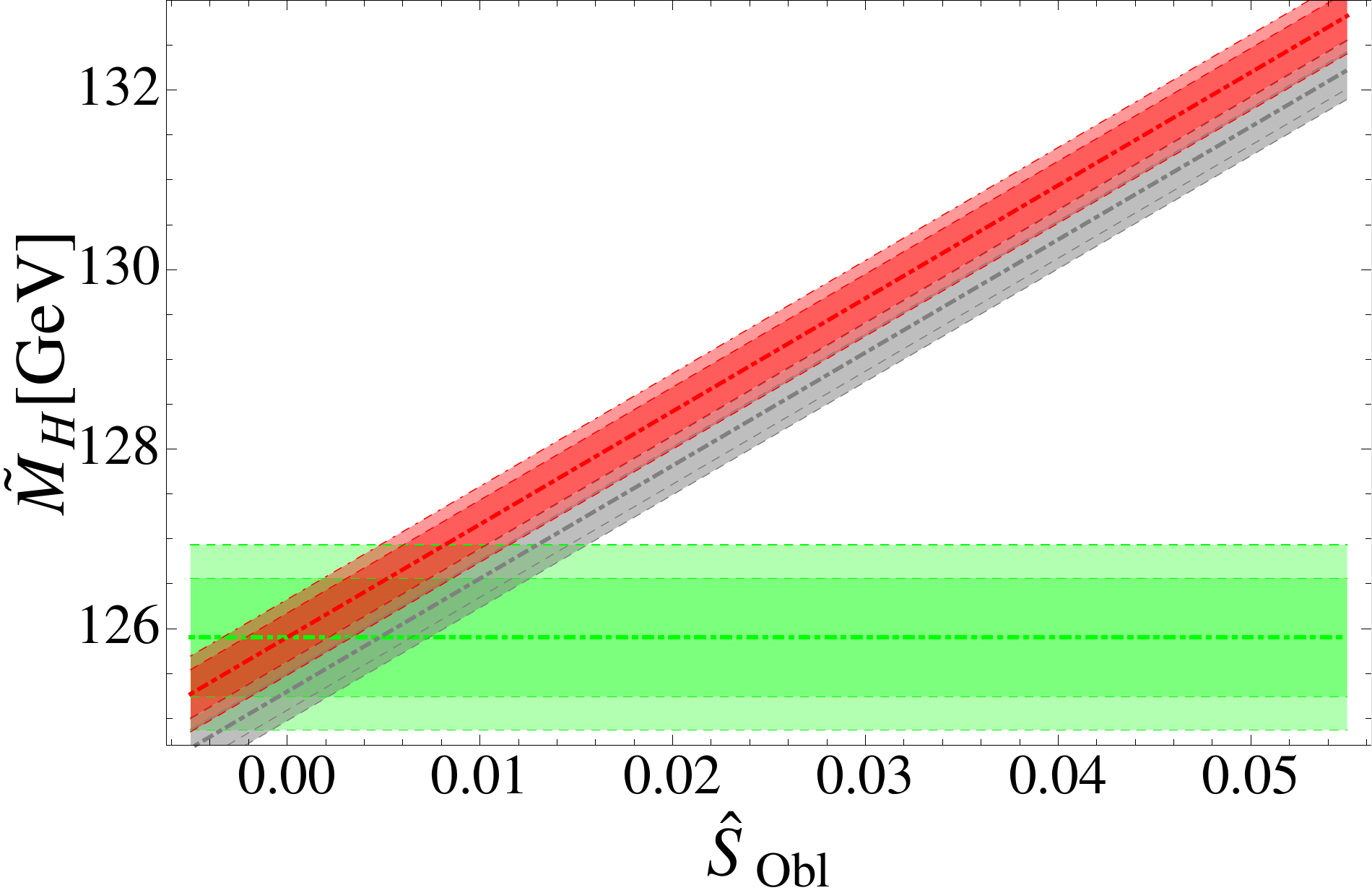} 
\caption{\label{fig:2} 90\% and 99\% CL contour plots for the rescaling of the effective Top and Higgs masses,  left and right figure respectively,  resulting from the ideal cancellation in the parameter $\hat S$ (red regions) and $\eps_3$ (gray regions). $\tilde m_t $ and $\tilde M_H  $ are the effective values calculated in custodial models by only  considering oblique contributions, $m_t = 173.4 \pm 1.0 \gev$ and $M_H = 125.9 \pm 0.4 \gev $ (green regions) are the physical values obtained by also considering direct corrections.}
\end{figure}

 To reproduce the correct physical spectrum in the Higgs and fermionic sectors the corresponding effective parameters $\tilde \lambda$ and $\tilde \lambda_f$ of custodial models with only oblique corrections must be rescaled as 
 $\tilde \lambda \rightarrow  { \lambda}{(1 + \hat S_{Obl})}^2$ and 
 $\tilde \lambda_f \rightarrow  {\lambda_f }{(1 + \hat S_{Obl})}$ when the ideal cancellation is considered. 
 The rescaling of the effective fermion and Higgs masses $\tilde m_f=  \frac{1}{\sqrt 2} \lambda_f (1 + \hat S_{Obl}) \vt $ and $\tilde M_H =  \sqrt{2 \lambda}(1 + \hat S_{Obl}) \vt$ with respects to their physical values is shown in FIG.\ref{fig:2}.  For instance, an universal contribution $\mathcal C_2$ implies that the effective value of the quark Top mass obtained by only considering the effect of oblique corrections must be higher to obtain, after considering the effect of the direction corrections to the EW scale, the experimental value $m_t$. This could compromise the benefit to introduce $\mathcal C_2$ to cure the $Z b \bar b$ coupling.

\begin{figure}[tbp]
\centering 
\hspace{1em}\includegraphics[width=0.6\textwidth]{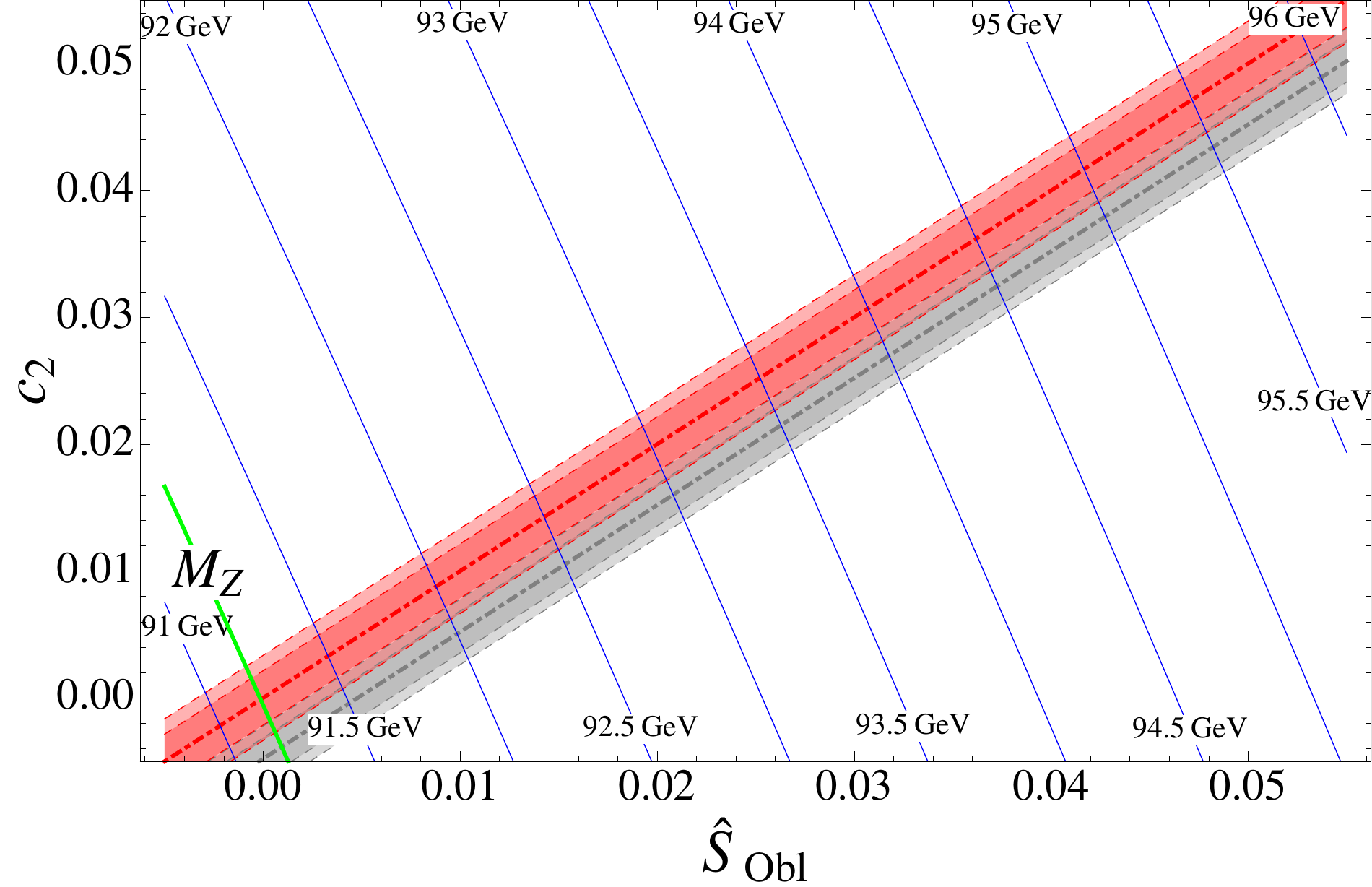}
\caption{\label{fig:3} 90\% and 99\% CL bounds for the rescaling of the effective $Z$ mass $\tilde M_Z  $ (blue lines) with respect the physical value  $M_Z = 91.188 \pm 0.002 \gev $ (green line) in terms of the oblique and direct corrections, $\hat S_{Obl}$ and $c_2$ respectively, to the $\hat S$ (red region) and $\eps_3$ (gray region) parameters.}
\end{figure}

The effect of the rescaling in terms of the oblique and direct parameters $\hat S_{Obl}$ and $c_2$ of the effective $W$ mass,  $\tilde M_W = \frac{\vt}{ 2} \sqrt{\frac{g^2}{(1-c_2)^2} + \gp^2 (1+ \hat S_{Obl})^2}$,  with respect to its physical value of $M_W$ is shown in FIG.\ref{fig:3}.  

Similarly, by assuming SM Higgs couplings with  the effective fields $\Wt^a_\mu$ and $\Bt_\mu$, the redefinitions (\ref{redef:S:canc}) implies the following modification with respect to the physical fields in the SM Higgs sector: $\frac{g^2}{2} (H^2 + 2 H \vt ) (1 + \hat S_{Obl})^2(W^+W^- + Z^2 / 2 c^2_\theta) $. Thus we have the following rescalings for the SM Higgs couplings $\tilde g_{H f f} \rightarrow g_{H f f} (1 + \hat S_{Obl})$,  $\tilde g_{H V V} \rightarrow g_{H V V} (1 + \hat S_{Obl})$,  $\tilde g_{H H V V} \rightarrow g_{H H V V} (1 + \hat S_{Obl})^2$ and $\tilde g_{H H H} \rightarrow g_{H H H} /(1 + \hat S_{Obl})$.
 
Notice however that in phenomenological models new physics typically contributes to the SM masses. These contributions may be compensated by the rescaling associated to the ideal cancellation. As exemplification we consider the case of a custodial Randall-Sundrum model with brane localized Higgs \cite{Casagrande:2010si,Casagrande:2008hr} (the phenomenology of this model is in fact investigated at higher order corrections of new physics). In this case the effective EW scale $\vt$, determined through the Fermi constant \cite{Casagrande:2008hr}, is related to the Higgs vacuum expectation value $v_H$ by the relation $\vt \simeq v_H \left[ 1 - \frac{\tilde M^2_W}{ 4 M^2_{KK}}(1 - \frac{1}{2L}) \right]$ where $v_H$ is the v.e.v. of the Higgs boson, $M_{KK}$ is the mass scale of the low-lying Kaluza-Klein excitations and $L = \ln\frac{ k }{ M_{KK}} $. The oblique contribution to the $S$ parameter is $\hat S_{Obl} = \frac{\gt^2}{8 M^2_{KK}} (1 - \frac{1}{L})$ --- the corrections to $T$ associated to new physics contributions to the Higgs sector will be investigate in future papers, see also sec.(\ref{theo:orig}). Hence, upon ideal cancellation, the contribution of direct corrections in the fermionic sector implies the following redefinition of the effective EW energy scale  $ v = \vt (1+\hat S_{Obl}) =  v_H \left[ 1 + \frac{\tilde M^2_W}{ 4 M^2_{KK}}(1 - \frac{3}{2L}) \right] $.

 In sec.(\ref{spurious}) we will extend the ideal cancellation to the full effective Lagrangian including the cancellation on the non-abelian and transverse terms. 
Indeed, the \emph{ideal} cancellation is an interesting ``effective symmetry'' of the SM, i.e. the SM Lagrangian can be defined modulo oblique and direct corrections. Thus the unwanted oblique and direct contributions of new physics can be separately greater than the actual experimental bounds on $S$.  

As well-known the ideal cancellation yields an extended parameter space to phenomenological models \cite{Anichini:1994xx,Cacciapaglia:2004rb,Casalbuoni:2005rs,Chivukula:2005cc}, allowing for a lower energy scale of new physics (especially in Randall-Sundrum models due to the warped factor).  In addition to this, provided that the new physics modes are sufficiently heavy to be eliminated in the EoM, the Lagrangian analysis of the ideal cancellation shows that this bound to new physics is further reduced by the redefinition of the effective EW energy scale $\vt$ with respect to $v$, described in FIG.(\ref{fig:1}).

\section{Ideal cancellation as ``effective symmetry''}\label{effective:sym}

So far we have investigated the ideal cancellation by considering only the transverse components of the gauge bosons. 
We now generalize our analysis to the longitudinal and non-abelian terms. We continue to assume that the custodial new physics sector is sufficiently heavy with respect to the SM particles to be neglected in the EoM. Assuming a SM Higgs mechanism,  the corrections of new physics are encoded in  custodial dimension-six operators in the gauge and fermionic sectors. We now infer the general form of these operators from considerations about the custodial symmetry and the elimination of the heavy modes from the EoM. In particular we find that the resulting operator $\hat {\mathcal O}_{WB}$ in the gauge sector coincides with $\mathcal O_{WB}$ in the transverse terms but it has a different form in the non-abelian terms. Thus, the general operator $\hat{\mathcal{O}}_{WB}$ gives the same contribution of $\mathcal O_{WB}$ to the $S$ parameter. Furthermore, contrarily to the operator $\mathcal O_{WB}$, the general operator $\hat {\mathcal O}_{WB}$ is exactly eliminated by the fermionic operator $\mathcal C_2$  upon ideal cancellation. That is, the cancellation occurs in the full effective Lagrangian (including non-abelian terms) giving back exactly the SM Lagrangian with rescaled EW energy.  

The operator $\mathcal {O}_{WB}$, (\ref{op:S:Obl}), commonly used in literature, e.g. \cite{BuchmŸller1986621,Han:2004az,Barbieri:2004qk,Fichet:2013ola}, as source of the oblique corrections to $S$ generates anomalous trilinear corrections of the type 
 $ \sim \hat S_{Obl} \gpt  \Wt^+_\mu \Wt^-_\nu \Bt^\munu$, \cite{Han:2004az}. 
By following the same line of the Lagrangian analysis in sec.(\ref{lagrang:analysis}), we find that the  transformation of $\Wt^3$ in (\ref{redef:S:canc}) eliminates these anomalous corrections.  However the redefinition of $\gt$ generates corrections to the non-abelian terms which are not cancelled by the condition of ideal cancellation (\ref{ideal:canc}). The residual corrections of $\mathcal {O}_{WB}$ in the trilinear and quartic vector boson couplings are respectively $g_{trilin} = (1+ \hat S_{Obl}) g^{SM}_{trilin} $ and $g_{quart} = (1+ \hat S^2_{Obl}) g^{SM}_{quart} $.  These residual corrections can be relevant in the analysis of the longitudinal components associated to the elastic scattering amplitude $W W \rightarrow W W$ and in turns in the unitarization of the theory. Moreover, by considering the corrections in the SM Higgs couplings described in the previous section, they imply a further enhancement of $H \rightarrow \gamma \gamma$ decay rate with respect to the enhanced $H \rightarrow V V$ decays whereas $H \rightarrow \bar f f$ is unchanged. This effect could be useful in interpreting possible excesses in the Higgs to vector bosons decay,  especially excesses in $H \rightarrow \gamma \gamma$.  These spurious corrections in the non-abelian terms are exactly eliminated by the ideal cancellation as soon as we consider the general custodial operator $\hat{\mathcal{O}}_{WB}$.

\subsection{Exact cancellation of all spurious corrections}\label{spurious}

Phenomenological models typically involve extended EW gauge symmetries such as in composite, little Higgs, moose, technicolor models, or  bulk gauge symmetries in extra-dimensional models. The low energy corrections to the gauge sector are obtained by eliminating from the EoM the related heavy modes  in terms of the effective SM fields $\Wt_\mu^a$ and $\Bt_\mu$ \footnote{Notice that in extra-dimensional phenomenology the holographic prescription is equivalent to the elimination of the KK modes in terms of the SM vector bosons, i.e. the so called source fields \cite{Casalbuoni:2007xn}. Indeed the holographic approach  is often used in extra-dimensional theories to derive the effective Lagrangian.}. In custodial models, if we denote the new physics gauge bosons  $\mathcal{V}_\mu^i  \in G \supseteq SU_L(2) \otimes SU_R(2)$, with generic coupling $g_{\mathcal V}$ and generators $\mathcal{T}^i$,   their elimination from the EoM leads to an effective solution of the following general $SU_D(2)$ custodial invariant  form 
\be\label{EOM:eff:cust}
g'' \mathcal{V}_\mu^i \mathcal{T}^i \leadsto h(p^2) \left(- \gt \Wt^a \frac{\sigma^a}{2} + \gpt \Bt \frac{\sigma^3}{2}\right) \in SU_D(2)\,, 
\ee
 where $h(p^2)$ is a  function encoding the new physics corrections.

By assuming direct couplings of the left-handed SM fermions with the new physics sector, in custodial models we have the following modification of the corresponding SM covariant derivative
  \be\label{cov:der:gauge:L}
  \tilde D_\mu \psi_L = \left[ \ptl_\mu - i \gt \Wt_\mu \frac{\sigma^a}{2} -  i \gpt \Bt_\mu \frac{B-L}{2} +  i c_2 \left(\gt \Wt_\mu \frac{\sigma^a}{2} - i \gpt \Bt_\mu \frac{\sigma^3}{2}\right ) \right] \psi_L
  \ee 
  The coefficient $c_2$ describes the overlap between the fermions and the new physics vector bosons. 
  By comparison with (\ref{S:dir:operator}) we see that actually (\ref{EOM:eff:cust}) correctly generates the direct operator $\mathcal C_2$ investigated in sec.(\ref{Direct:corrections}).    
  
 Similarly we now derive the general form of the custodial operator $\hat {\mathcal O}_{WB}$ in the gauge sector. In the field strength,  the elimination of the heavy modes from the EoM induces the following modification of the SM covariant derivative 
 \be\label{cov:der:gauge}
\tilde D_\mu =  \ptl_\mu - i \gt \Wt_\mu \frac{\sigma^a}{2}   - i \gpt \Bt_\mu  + i \frac{\hat S_{Obl}}{1 + \hat S_{Obl}} \left (\gt \Wt^a_\mu \frac{\sigma^a}{2} -  \gpt \Bt_\mu \frac{\sigma^3}{2} \right) \,,
  \ee  
 where $\hat S_{Obl}$ encodes the coupling of the heavy vector bosons with the SM  $\Wt^a$ and $\Bt$.

The field strength associate  to the effective custodial theory is therefore directly inferred from the modified covariant derivative (\ref{cov:der:gauge}) as 
\be
[\tilde D_\mu, \tilde D_\nu] = - i g \mathcal{F}^a_\munu  \frac{\sigma^a}{2} - i \gpt \Bt_\munu \,. \label{field:stre}
\ee 
where, for convenience, we have parameterized $\gt \rightarrow g (1 + \hat S_{Obl})$ --- this avoids a further normalization of $\Wt$. Notice that upon the condition of ideal cancellation this choice corresponds to the reparametrization in (\ref{redef:S:canc}). 

Thus the effective $\Wt$ field strength encoding custodial new physics corrections in the gauge sector has the form
\be
 \mathcal{F}^a_\munu   \frac{\sigma^a}{2} = F^a_\munu(\Wt)\frac{\sigma^a}{2} + \frac{\hat S_{Obl}}{1 + \hat S_{Obl}} \Delta{\mathcal G}_\munu^a   \frac{\sigma^a}{2} + \hat S_{Obl}\frac{ \gpt}{\gt}  \frac{\sigma^3}{2} \Bt_\munu 
 \ee
 where
 \begin{eqnarray}
 \Delta{\mathcal G}_\munu^\pm  &=&  - i \gt (\Wt^-_\mu \Wt_\nu^+ - \Wt^-_\nu \Wt_\mu^+ )  \nn \\
 \Delta{\mathcal G}_\munu^3  &=&  \mp i \gt (\Wt_\nu^3 \Wt_\mu^\pm - \Wt^3_\mu \Wt^\pm_\nu) \mp i   \gpt (\Bt_\nu \Wt_\mu^\pm - \Bt_\mu \Wt^\pm_\nu) 
 \end{eqnarray} 
 The explicit derivation of these non-abelian custodial corrections in the gauge kinetic terms is given in \cite{Dolce:PhDThesis} for the specific cases of a moose and an extra-dimensional custodial model.
 Thus, (upon unphysical normalizations of $\Bt$ and $\gpt$) the general operator $\hat{\mathcal{O}}_{WB}$ encoding custodial corrections in the kinetic gauge terms is 
\begin{eqnarray}\label{correct:oblique:operator}
\hat{\mathcal O}_{WB} &=&  {\mathcal O}_{WB} - \frac 1 2  \frac{\hat S_{Obl} }{1 + \hat S_{Obl}} \left (  \hat S_{Obl} \frac{\gpt}{\gt}  \Delta{\mathcal G}_\munu^3 \Bt^\munu  +  2 \text{Tr}[ F^a_\munu (\Wt) \Delta{\mathcal G}^{a \munu}] \right. \nn \\ &+& \left.  \frac{  \hat S_{Obl} }{1 + \hat S_{Obl}}  \text{Tr}[\Delta{\mathcal G}_\munu^a \Delta{\mathcal G}^{a \munu} ] \right)
\end{eqnarray}
 Its bilinear terms are equivalent to $\mathcal O_{WB}$  so that both $\mathcal O_{WB}$ and  $\hat {\mathcal O}_{WB}$  generate the same oblique correction to $ S$. This confirms our model independent derivation of custodial corrections in terms of the elimination of the heavy fields from the EoM. At the same time this shows that in general the custodial corrections to the gauge kinetic terms differ in the non-abelian terms from the commonly used operator $\mathcal O_{WB}$.

The general derivation of the custodial corrections in terms of redefinitions of the covariant derivatives  is particularly convenient.  In this way it is in fact straightforward to see from the covariant derivatives (\ref{cov:der:gauge:L}) and (\ref{cov:der:gauge}) that the redefinition of  $\Wt^3$ and $\gt$  given (\ref{redef:S:canc}) exactly cancel all the new physics corrections in the gauge invariant terms of the effective Lagrangian. The only remaining corrections are in the spontaneously broken gauge terms such as the vector boson mass term. In particular this implies the redefintion of the EW energy scale (\ref{vev:rescaling:exact}).  
Furthermore, from  (\ref{cov:der:gauge:L}) and (\ref{cov:der:gauge}) we immediately see that the condition of ideal cancellations  (\ref{ideal:canc}) means, in terms of gauge invariance, that the coefficients of the extra gauge terms must be exactly the same in the two covariant derivatives. We will discuss further this interesting aspect in the next section. 

Thus, the condition of ideal cancellation exactly eliminates all the new physics contributions in the effective Lagrangian giving back the SM Lagrangian with a possible redefinition of the EW scale, and thus of the parameters in the Higgs and Yukawa sectors,
\begin{eqnarray}
\boxed{\boxed{\call^{Eff}_{Tot}(\tilde{v}) =  \call^{SM}(\vt) + \hat {\mathcal O}_{WB} + \mathcal C_2 \equiv \call^{SM}({v})}}\,.
\end{eqnarray}
Upon the condition of ideal cancellation and rescaling of the effective EW energy $\vt$, the gauge and fermionic sectors of SM Lagrangian can be defined modulo custodial corrections, namely $\hat{\mathcal O}_{WB}$ and $\mathcal C_2$. We conclude that, on the basis of general principles, the ideal cancellation represents a new ``\emph{effective symmetry}'' of the SM Lagrangian, i.e. a symmetry valid in the low energy approximation and protecting the custodial effective Lagrangian from corrections with respect to the SM lagrangian.

\subsection{On the theoretical meaning of the ideal cancellation}\label{theo:orig}

It is worthwhile investigating the possible theoretical origin of the ``effective symmetry'' associated to the ideal cancellation. Here we study the origin of the ideal cancellation in terms of gauge invariance. Though the ideal cancellation is a very model dependent issue, our considerations about the modification of the covariant derivative allows us an interesting model independent analysis. 

In extra-dimensional (including gauge-Higgs unification) or composite and moose models, the direct couplings of fermions to new physics can be associated to delocalizations of the fermions through a Wilson line of the fifth component of the bulk gauge fields or of the direct product of $\sigma$ model scalar fields, e.g. \cite{Casalbuoni:2005rs,Dolce:PhDThesis}. Through the eliminations of the new physics d.o.f. from the EoM in the derivation of the effective theories, such a delocalization can be described in terms of the SM fields.
 For instance the direct correction $\mathcal C_2$, i.e. the covariant derivative (\ref{cov:der:gauge:L}),  can be obtained from the effective Wilson line \be\psi_L(x) \rightarrow e^{-i \pi_{Dir}^a(x) \frac{\sigma^a}{2}} \psi_L(x)\ee with the custodial invariant contribution 
\be 
\pi_{Dir}^a(x) \frac{\sigma^a}{2} = - c_2 \int^x dx'^\mu (\gt \Wt_\mu \frac{\sigma^a}{2}- \gpt \Bt_\mu \frac{\sigma^3}{2} ) ~~~~~~\Leftrightarrow ~~~~~~\mathcal C_2\,.
\ee 

 Similarly, the oblique corrections in the gauge sector (\ref{cov:der:gauge}), namely the operator $\hat{\mathcal O}_{WB}$, can be associated to an analogous custodial effective Wilson line with modulated phase  
\be
\pi_{Obl}^a(x) \frac{\sigma^a}{2} =  -  \frac{  \hat S_{Obl} }{1 + \hat S_{Obl}}   \int^{x} dx^{\prime \mu} (\gt \Wt_\mu \frac{\sigma^a}{2}- \gpt \Bt_\mu \frac{\sigma^3}{2} ) ~~~~~~  \Leftrightarrow ~~~~~~\hat{\mathcal O}_{WB}\,. 
\ee 
Even though the coefficients of the Wilson lines responsible for the oblique and direct corrections are in general different in effective models, we see that the condition of ideal cancellation is related to the effective gauge invariance of the theory. For instance, gauge invariance implies that the SM couplings are such that $g_{Vff} = g_{VVV}$:  the gauge couplings in the fermionic covariant derivative and in the field strength must be the same. In terms of the coefficients of the oblique and direct corrections this implies the ideal cancellation condition (\ref{ideal:canc}). Roughly speaking the ideal cancellation can be regarded as the gauging of an extra custodial group $SU_D(2)$ with  reduced couplings by factors $\frac{  \hat S_{Obl} }{1 + \hat S_{Obl}}$, or equivalently $c_2$, with respect to the SM EW couplings in both the gauge and fermion sectors.  

In case of ideal cancellation, and as long as the new physics d.o.f. are sufficiently heavy, the pure custodial corrections to the gauge and fermionic sector of the SM can be in principle much higher that the actual experimental bounds on $S$.  This can imply sizable redefinitions of the effective parameters of the models, such as EW scale $\vt$, couplings and mass spectrum, and thus of the allowed scale for new physics. 

This paper is exclusively dedicated to the study of the $S$ parameter in purely custodial effective theories. Nevertheless, as we will see in future papers,  our analysis can be extended to the other EW parameters by considering the generalization of the ideal cancellation of other possible operators among the different sectors of the SM lagrangian \cite{Han:2004az}.  

Notice that if we allow direct coupling of the Higgs field to the custodial new physics, the covariant derivative of the Higgs boson turns out to have a gauge structure similar to those of the field strength and of the left-handed fermions, (\ref{cov:der:gauge:L}) and (\ref{cov:der:gauge}). The coefficient $c_U$ in front of the non-standard contribution would be given by the overlap between the Higgs and the vector boson profiles.  In the Higgs sector, the dimension six operator responsible for the custodial corrections is in fact $  |U^\dagger D_\mu U|^2  \leadsto c_U \left (\gt \Wt^a_\mu \frac{\sigma^a}{2} -  \gpt \Bt_\mu \frac{\sigma^3}{2} \right)^2$. By generalizing the concept of ideal cancellation to the Higgs sector we can assume ideal direct coupling or delocalization of the Higgs field such that this coefficient is the same of the other sectors, \cite{Quiros:2013yaa}. Hence the redefinitions (\ref{redef:S:canc}) exactly reabsorb the possible corrections in the Higgs sector, obtaining the SM Lagrangian with possible rescaling of the EW energy.  An ideal delocalization of the Higgs profile could also play an interesting role in protecting the $T$ parameter from corrections associated to a non-SM Higgs, as suggested by the analysis given in the case of extra-dimensional theories \cite{Quiros:2013yaa}.

Finally we want to mention that, in the context of the theory summarized in \cite{Dolce:cycles}, this paper represents a preparatory study for a possible interpretation of the AdS/QCD correspondence in terms of a ``virtual extra-dimension" \cite{Dolce:AdSCFT,Dolce:ICHEP2012} and that the considerations about the covariant derivatives can be interpreted in terms of \cite{Dolce:tune}.

\addcontentsline{toc}{section}{Conclusions}
\section*{Conclusions}

The discovery of the light SM-like Higgs-boson at LHC demands for new mechanics to make the EW physics insensible to new physics. In this paper we show that large oblique contributions to the $S$ parameter can be exactly eliminated from the SM Lagrangian by introducing direct custodial corrections in the fermionic sector \cite{Anichini:1994xx,Cacciapaglia:2004rb,Casalbuoni:2005rs,Chivukula:2005xm,Casalbuoni:2007xn}. The main fermionic operator investigated in this paper is, for instance, introduced in model building to cure tensions in the $Z b \bar b$ coupling \cite{Agashe:2006at,Casagrande:2008hr,Cui:2009dv}. The ideal cancellation among oblique and direct corrections turns out to be associated to possible sizable redefinition of the effective EW energy scale with respect to models with only oblique corrections, allowing for an extended parameter space for custodial new physics. In turns this implies redefinitions of the vector boson masses, Yukawa parameters, Higgs mass and couplings. From the general principle of custodial invariance and from the elimination of the heavy modes from the EoM, we have inferred  the exact form $ \hat {\mathcal O}_{WB} $ of the custodial corrections of new physics in the gauge sector. Upon ideal cancellation such custodial corrections in the gauge sector can be exactly eliminated by direct corrections in the fermionic sector. Besides  the elimination of corrections in the transverse components of the gauge bosons, this also eliminates the corrections in the longitudinal and non-abelian terms. Thus, as long as the new vectorial modes can be neglected in the effective theory, the SM Lagrangian turns out to be defined modulo custodial corrections in the gauge and fermionic sector (provided possible rescalings in the Higgs and Yukawa sector). Similarly to the custodial symmetry protecting the effective theory from corrections to the $T$ and $U$ parameters, the ideal cancellation can be therefore thought of as an "effective symmetry" protecting the SM from custodial corrections. It is therefore important to investigate the possible theoretical origin of such an ``effective symmetry''. In general, this implies a fine-tuning between the couplings of new physics with the SM gauge and fermionic sectors. We have found, however, that such a fine-tuning has a simple justification in terms of the universality, among the gauge and fermionic sectors, of the effective gauge structure associated to the new physics contributions.   
\acknowledgments
I would like to thank Prof. A. Pomarol,  Prof. D. Dominici, Prof. S. de Curtis, Prof. M. Neubert and Prof. T. Gherghetta for fruitful discussions. 
This work was partially supported by the Angelo della Riccia Foundations. 
\appendix

\section{Ideal cancellation without rescaling }

It is possible to study other custodial fermionic operators with negative contributions to the $S$ parameter, so that it is possible to define new conditions of ideal cancellation that exactly reduces the effective Lagrangian to the SM one. As in the case studied in the text, the cancellation in the non-abelian terms is possible by introducing the general custodial operator $\hat {\mathcal{O}}_{WB}$. Here we will investigate a universal custodial operator in the fermionic sector acting on the hypercharge. In this case however the ideal cancellation will not imply a rescaling of the effective EW scale with respect to its physical value.

The correction that we want to investigate is 
\begin{eqnarray}
\mathcal C_Y =  c_Y i \bar\psi  \gamma_\mu   \gpt \Bt^\mu  \left[-\frac{\sigma^3}{2} - \frac{(B-L)}{2}\right] \psi  
  \label{S:dir:right:operator}
\end{eqnarray}
with positive defined $c_Y$ and $\psi = \psi_L + \psi_R$. 
By assuming an extended electroweak gauge symmetry $SU_L(2)\times SU_R(2) \times U_{B-L}(1)$ where the EW symmetry is recovered through the breaking $SU_R(2) \times U_X(1) \rightarrow U_Y(1)$, this correction can be associated to the operator
$
\mathcal C_Y =  \Tr [\bar\psi \gamma_\mu  (i D^\mu \Sigma_Y) \Sigma_Y^\dagger \psi] 
$
where $X$ transforms as $ \Sigma_{Y} \rightarrow R \Sigma_{Y} X$ where $X \in U_{Y}(1)$ and $T_L^3 = - T_R^3$.  
The standard analysis of the EW constrains yields 
$  \hat S = \hat S_{Obl} - c_Y$, $U \sim 0$ and $T \sim 0$, 
so that the ideal cancellation of the oblique and direct corrections is $
   c_Y  \sim \hat S_{Obl} $:
\be
\hat S \sim 0~,~~ \hat U \sim 0~, ~~ \hat T \sim 0\,.
\ee 
 
This result can be checked at the Lagrangian level. The effective Lagrangian containing both $\hat {\mathcal O}_{WB}$ and $\mathcal C_Y$ is reduced to the SM Lagrangian by imposing 
the following redefinition (upon unphysical redefinition of $\Bt$ and $\gpt$ normalizing the kinetic term) of the effective fields and couplings  
$\Wt^a_\mu \frac{\sigma^a}{2} \rightarrow  W^a_\mu \frac{\sigma^a}{2} - \hat S_{Obl} \frac{\gpt}{\gt}  B_\mu \frac{\sigma^3}{2}$, $
\gpt \rightarrow   \frac{ \gp }{1 + c_Y}$. 

In this way, by following the line of the Lagrangian analysis in sec.(\ref{lagrang:analysis}), we find that the ideal cancellation is given by the condition 
\be
c_Y \equiv S_{Obl}
\ee
so that the effective Lagrangian is reduced to the SM Lagrangian
 \be
 \call^{Eff}_{Tot} =   \call^{SM} + \hat {\mathcal O}_{WB} + \mathcal C_Y \equiv \call^{SM}
 \ee 
 
Notice  that in this case there is no rescaling of $\vt$. In fact the contribution coming from the redefinition of $\Wt^a$ in the vector boson mass term is completely reabsorbed by the redefinition of $\gpt$.
Thus we have obtained another instance of ``effective symmetry'' of the SM lagrangian: upon ideal cancellation the SM Lagrangian can be defined modulo operators $\hat {\mathcal O}_{WB}$ and $\mathcal C_Y$ in the gauge and fermionic sector, respectively.

\bibliographystyle{JHEP}
\bibliography{comp3+1}

\providecommand{\href}[2]{#2}\begingroup\raggedright\begin{thebibliography}{10}

\bibitem{Beringer:1900zz}
{\bf Particle Data Group} Collaboration, J.~Beringer {\em et~al.}, {\it {Review
  of Particle Physics (RPP)}},  {\em Phys.Rev.} {\bf D86} (2012) 010001.

\bibitem{Peskin:1991sw}
M.~E. Peskin and T.~Takeuchi, {\it {Estimation of oblique electroweak
  corrections}},  {\em Phys. Rev.} {\bf D46} (1992) 381--409.

\bibitem{Altarelli:1993sz}
G.~Altarelli, R.~Barbieri, and F.~Caravaglios, {\it {Nonstandard analysis of
  electroweak precision data}},  {\em Nucl. Phys.} {\bf B405} (1993) 3--23.

\bibitem{Agashe:2007mc}
K.~Agashe, C.~Csaki, C.~Grojean, and M.~Reece, {\it {The S-parameter in
  holographic technicolor models}},  {\em JHEP} {\bf 0712} (2007) 003,
  [\href{http://arxiv.org/abs/0704.1821}{{\tt arXiv:0704.1821}}].

\bibitem{SekharChivukula:2009if}
R.~Sekhar~Chivukula, S.~Di~Chiara, R.~Foadi, and E.~H. Simmons, {\it {The
  Limits of Custodial Symmetry}},  {\em Phys.Rev.} {\bf D80} (2009) 095001,
  [\href{http://arxiv.org/abs/0908.1079}{{\tt arXiv:0908.1079}}].

\bibitem{Mintakevich:2009wz}
O.~Mintakevich and J.~Sonnenschein, {\it {Holographic technicolor models and
  their S-parameter}},  {\em JHEP} {\bf 0907} (2009) 032,
  [\href{http://arxiv.org/abs/0905.3284}{{\tt arXiv:0905.3284}}].

\bibitem{Cort:2013foa}
L.~Cort, M.~Garcia, and M.~Quiros, {\it {Supersymmetric Custodial Triplets}},
  {\em Phys.Rev.} {\bf D88} (2013) 075010,
  [\href{http://arxiv.org/abs/1308.4025}{{\tt arXiv:1308.4025}}].

\bibitem{Carena:2014ria}
M.~Carena, L.~Da~Rold, and E.~Ponton, {\it {Minimal Composite Higgs Models at
  the LHC}},  \href{http://arxiv.org/abs/1402.2987}{{\tt arXiv:1402.2987}}.

\bibitem{Barbieri:2004qk}
R.~Barbieri, A.~Pomarol, R.~Rattazzi, and A.~Strumia, {\it {Electroweak
  symmetry breaking after LEP1 and LEP2}},  {\em Nucl. Phys.} {\bf B703} (2004)
  127--146, [\href{http://arxiv.org/abs/hep-ph/0405040}{{\tt hep-ph/0405040}}].

\bibitem{Casagrande:2010si}
S.~Casagrande, F.~Goertz, U.~Haisch, M.~Neubert, and T.~Pfoh, {\it {The
  Custodial Randall-Sundrum Model: From Precision Tests to Higgs Physics}},
  {\em JHEP} {\bf 1009} (2010) 014, [\href{http://arxiv.org/abs/1005.4315}{{\tt
  arXiv:1005.4315}}].

\bibitem{BuchmŸller1986621}
W.~BuchmŸller and D.~Wyler, {\it Effective lagrangian analysis of new
  interactions and flavour conservation},  {\em Nuclear Physics B} {\bf 268}
  (1986), no.~3Ð4 621 -- 653.

\bibitem{Han:2004az}
Z.~Han and W.~Skiba, {\it {Effective theory analysis of precision electroweak
  data}},  {\em Phys.Rev.} {\bf D71} (2005) 075009,
  [\href{http://arxiv.org/abs/hep-ph/0412166}{{\tt hep-ph/0412166}}].

\bibitem{Agashe:2006at}
K.~Agashe, R.~Contino, L.~Da~Rold, and A.~Pomarol, {\it {A custodial symmetry
  for Z b anti-b}},  {\em Phys. Lett.} {\bf B641} (2006) 62--66,
  [\href{http://arxiv.org/abs/hep-ph/0605341}{{\tt hep-ph/0605341}}].

\bibitem{Anichini:1994xx}
L.~Anichini, R.~Casalbuoni, and S.~De~Curtis, {\it Low-energy effective
  lagrangian of the bess model},  {\em Phys. Lett.} {\bf B348} (1995) 521--529,
  [\href{http://arxiv.org/abs/hep-ph/9410377}{{\tt hep-ph/9410377}}].

\bibitem{Cacciapaglia:2004rb}
G.~Cacciapaglia, C.~Csaki, C.~Grojean, and J.~Terning, {\it Curing the ills of
  higgsless models: The s parameter and unitarity},  {\em Phys. Rev.} {\bf D71}
  (2005) 035015, [\href{http://arxiv.org/abs/hep-ph/0409126}{{\tt
  hep-ph/0409126}}].

\bibitem{Casalbuoni:2005rs}
R.~Casalbuoni, S.~De~Curtis, D.~Dolce, and D.~Dominici, {\it Playing with
  fermion couplings in higgsless models},  {\em Phys. Rev.} {\bf D71} (2005)
  075015, [\href{http://arxiv.org/abs/hep-ph/0502209}{{\tt hep-ph/0502209}}].

\bibitem{Chivukula:2005xm}
R.~S. Chivukula, E.~H. Simmons, H.-J. He, M.~Kurachi, and M.~Tanabashi, {\it
  {Ideal fermion delocalization in Higgsless models}},  {\em Phys.Rev.} {\bf
  D72} (2005) 015008, [\href{http://arxiv.org/abs/hep-ph/0504114}{{\tt
  hep-ph/0504114}}].

\bibitem{Casalbuoni:2007xn}
R.~Casalbuoni, S.~De~Curtis, D.~Dominici, and D.~Dolce, {\it {Holographic
  approach to a minimal Higgsless model}},  {\em JHEP} {\bf 08} (2007) 053,
  [\href{http://arxiv.org/abs/0705.2510}{{\tt arXiv:0705.2510}}].

\bibitem{Quiros:2013yaa}
M.~Quiros, {\it {Higgs Bosons in Extra Dimensions}},
  \href{http://arxiv.org/abs/1311.2824}{{\tt arXiv:1311.2824}}.

\bibitem{Chivukula:2005cc}
R.~S. Chivukula, E.~H. Simmons, H.-J. He, M.~Kurachi, and M.~Tanabashi, {\it
  {Ideal fermion delocalization in five dimensional gauge theories}},  {\em
  Phys.Rev.} {\bf D72} (2005) 095013,
  [\href{http://arxiv.org/abs/hep-ph/0509110}{{\tt hep-ph/0509110}}].

\bibitem{Accomando:2010fz}
E.~Accomando, A.~Belyaev, L.~Fedeli, S.~F. King, and
  C.~Shepherd-Themistocleous, {\it {Z' physics with early LHC data}},  {\em
  Phys.Rev.} {\bf D83} (2011) 075012,
  [\href{http://arxiv.org/abs/1010.6058}{{\tt arXiv:1010.6058}}].

\bibitem{Casagrande:2008hr}
S.~Casagrande, F.~Goertz, U.~Haisch, M.~Neubert, and T.~Pfoh, {\it {Flavor
  Physics in the Randall-Sundrum Model: I. Theoretical Setup and Electroweak
  Precision Tests}},  {\em JHEP} {\bf 10} (2008) 094,
  [\href{http://arxiv.org/abs/0807.4937}{{\tt arXiv:0807.4937}}].

\bibitem{Cui:2009dv}
Y.~Cui, T.~Gherghetta, and J.~D. Wells, {\it {Emergent Electroweak Symmetry
  Breaking with Composite W, Z Bosons}},  {\em JHEP} {\bf 0911} (2009) 080,
  [\href{http://arxiv.org/abs/0907.0906}{{\tt arXiv:0907.0906}}].

\bibitem{Fichet:2013ola}
S.~Fichet and G.~von Gersdorff, {\it {Anomalous gauge couplings from composite
  Higgs and warped extra dimensions}},
  \href{http://arxiv.org/abs/1311.6815}{{\tt arXiv:1311.6815}}.

\bibitem{Contino:2010rs}
R.~Contino, {\it {The Higgs as a Composite Nambu-Goldstone Boson}},
  \href{http://arxiv.org/abs/1005.4269}{{\tt arXiv:1005.4269}}.

\bibitem{Arkani-Hamed:2001ca}
N.~Arkani-Hamed, A.~G. Cohen, and H.~Georgi, {\it (de)constructing dimensions},
   {\em Phys. Rev. Lett.} {\bf 86} (2001) 4757--4761,
  [\href{http://arxiv.org/abs/hep-th/0104005}{{\tt hep-th/0104005}}].

\bibitem{Accomando:2012yg}
E.~Accomando, L.~Fedeli, S.~Moretti, S.~De~Curtis, and D.~Dominici, {\it
  {Charged di-boson production at the LHC in a 4-site model with a composite
  Higgs boson}},  {\em Phys.Rev.} {\bf D86} (2012) 115006,
  [\href{http://arxiv.org/abs/1208.0268}{{\tt arXiv:1208.0268}}].

\bibitem{Dolce:PhDThesis}
D.~Dolce, {\it {Higgsless extensions of the Standard Model of Electroweak
  Interacions}},  \href{http://arxiv.org/abs/D.D. page on
  www.ResearchGate.net}{{\tt D.D. page on www.ResearchGate.net}}. PhD thesis in
  Theoretical Physics, Florence University (2007).

\bibitem{Dolce:cycles}
D.~Dolce, {\it {Elementary spacetime cycles}},  {\em Europhys. Lett. ,} {\bf
  102} (2013) 31002, [\href{http://arxiv.org/abs/1305.2802}{{\tt
  arXiv:1305.2802}}].

\bibitem{Dolce:AdSCFT}
D.~Dolce, {\it {Classical geometry to quantum behavior correspondence in a
  Virtual Extra Dimension}},  {\em Annals Phys.} {\bf 327} (2012) 2354--2387,
  [\href{http://arxiv.org/abs/1110.0316}{{\tt arXiv:1110.0316}}].

\bibitem{Dolce:ICHEP2012}
D.~Dolce, {\it {AdS/CFT as classical to quantum correspondence in a Virtual
  Extra Dimension}},  {\em PoS} {\bf ICHEP2012} (2013) 478,
  [\href{http://arxiv.org/abs/1309.2646}{{\tt arXiv:1309.2646}}].

\bibitem{Dolce:tune}
D.~Dolce, {\it {Gauge Interaction as Periodicity Modulation}},  {\em Annals
  Phys.} {\bf 327} (2012) 1562--1592,
  [\href{http://arxiv.org/abs/1110.0315}{{\tt arXiv:1110.0315}}].

\end{thebibliography}\endgroup

\end{document}